# Secure and Privacy-Preserving Data Aggregation Protocols for Wireless Sensor Networks

Jaydip Sen
*Innovation Lab, Tata Consultancy Services Ltd.*
*India*

## 1. Introduction

In recent years, wireless sensor networks (WSNs) have drawn considerable attention from the research community on issues ranging from theoretical research to practical applications. Special characteristics of WSNs, such as resource constraints on energy and computational power and security have been well-defined and widely studied (Akyildiz et al., 2002; Sen, 2009). What has received less attention, however, is the critical privacy concern on information being collected, transmitted, and analyzed in a WSN. Such private and sensitive information may include payload data collected by sensors and transmitted through the network to a centralized data processing server. For example, a patient's blood pressure, sugar level and other vital signs are usually of critical privacy concern when monitored by a medical WSN which transmits the data to a remote hospital or doctor's office. Privacy concerns may also arise beyond data content and may focus on context information such as the location of a sensor initiating data communication. Effective countermeasure against the disclosure of both data and context-oriented private information is an indispensable prerequisite for deployment of WSNs in real-world applications (Sen, 2010a; Bandyopadhyay & Sen, 2011).

Privacy protection has been extensively studied in various fields such as wired and wireless networking, databases and data mining. However, the following inherent features of WSNs introduce unique challenges for privacy preservation of data and prevent the existing techniques from being directly implemented in these networks.

- *Uncontrollable environment*: sensors may have to be deployed in an environment that is uncontrollable by the defender, such as a battlefield, enabling an adversary to launch physical attacks to capture sensor nodes or deploy counterfeit ones. As a result, an adversary may retrieve private keys used for secure communication and decrypt any communication eavesdropped by the adversary.
- *Sensor-node resource constraints*: battery-powered sensor nodes generally have severe constraints on their ability to store, process, and transmit the sensed data. As a result, the computational complexity and resource consumption of public-key ciphers is usually considered unsuitable for WSNs.

- *Topological constraints*: the limited communication range of sensor nodes in a WSN requires multiple hops in order to transmit data from the source to the base station. Such a multi-hop scheme demands different nodes to take diverse traffic loads. In particular, a node closer to the base station (i.e., data collecting and processing server) has to relay data from nodes further away from base station in addition to transmitting its own generated data, leading to higher transmission rate. Such an unbalanced network traffic pattern brings significant challenges to the protection of context-oriented privacy information. Particularly, if an adversary has the ability to carry out a global traffic analysis, observing the traffic patterns of different nodes over the whole network, it can easily identify the sink and compromise context privacy, or even manipulate the sink node to impede the proper functioning of the WSN.

The unique challenges for privacy preservation in WSNs call for development of effective privacy-preserving techniques. Supporting efficient in-network data aggregation while preserving data privacy has emerged as an important requirement in numerous wireless sensor network applications (Acharya et al., 2005; Castelluccia et al., 2009; Girao et al., 2005; He et al., 2007; Westhoff et al., 2006). As a key approach to fulfilling this requirement of private data aggregation, *concealed data aggregation* (CDA) schemes have been proposed in which multiple source nodes send encrypted data to a sink along a *converge-cast tree* with aggregation of cipher-text being performed over the route (Acharya et al., 2005; Armknecht et al., 2008; Castelluccia et al., 2009; Girao et al., 2005; Peter et al., 2010; Westhoff et al., 2006).

He et al. have proposed a *cluster-based private data aggregation* (CPDA) scheme in which the sensor nodes are randomly distributed into clusters (He et al., 2007). The cluster leaders carry out aggregation of the data received from the cluster member nodes. The data communication is secured by using a shared key between each pair of communicating nodes for the purpose of encryption. The aggregate function leverages algebraic properties of the polynomials to compute the desired aggregate value in a cluster. While the aggregation is carried out at the aggregator node in each cluster, it is guaranteed that no individual node gets to know the sensitive private values of other nodes in the cluster. The intermediate aggregate value in each cluster is further aggregated along the routing tree as the data packets move to the sink node. The privacy goal of the scheme is two-fold. First, the privacy of data has to be guaranteed end-to-end. While only the sink could learn about the final aggregation result, each node will have information of its own data and does not have any information about the data of other nodes. Second, to reduce the communication overhead, the data from different source nodes have to be efficiently combined at the intermediate nodes along the path. Nevertheless, these intermediate nodes should not learn any information about the individual nodes' data. The authors of the CPDA scheme have presented performance results of the protocol to demonstrate the efficiency and security of the protocol. The CPDA protocol has become quite popular, and to the best of our knowledge, there has been no identified vulnerability of the protocol published in the literature so far. In this chapter, we first demonstrate a security loophole in the CPDA protocol and then proceed to show how the protocol can be made more secure and efficient.

Some WSN application may not require privacy of the individual sensor data. Instead, the data aggregation scheme may need high level of security so that no malicious node should be able to introduce any fake data during the execution of the aggregation process. This requirement introduces the need for design of secure aggregation protocols for WSNs.

Keeping this requirement in mind, we also present a secure and robust aggregation protocol for WSNs where aggregation algorithm does not preserve the privacy of the individual sensor data but guarantees high level of security in the aggregation process so that a potential malicious insider node cannot inject false data during the aggregation process.

The rest of this chapter is organized as follows. Section 2 provides a brief background discussion on the CPDA scheme. In Section 3, we present a cryptanalysis on CPDA and demonstrate a security vulnerability of the scheme. In Section 4, we present some design modifications of the CPDA scheme. Section 4.1 presents an efficient way to compute the aggregation operation so as to make CPDA more efficient. Section 4.2 briefly discusses how the identified security vulnerability can be addressed. Section 5 presents a comparative analysis of the overhead of the original CPDA protocol and its proposed modified version. Section 5.1 provides a comparison of the communication overheads in the network, and Section 5.2 provides an analysis of the computational overheads in the sensor nodes in the sensor nodes. Section 6 discusses the importance of security in designing aggregation schemes for WSNs. Section 7 presents some related work in the field of secure aggregation protocols in WSNs. In Section 8, a secure aggregation algorithm for WSNs is proposed. Section 9 presents some simulation results to evaluate the performance of the proposed secure aggregation protocol. Section 10 concludes the chapter while highlighting some future directions of research in privacy and security in WSNs.

## 2. The CPDA scheme for data aggregation in WSNs

The basic idea of CPDA is to introduce noise to the raw data sensed by the sensor nodes in a WSN, such that an aggregator can obtain accurate aggregated information but not individual sensor data (He et al., 2007). This is similar to the *data perturbation* approach extensively used in privacy-preserving data mining. However, unlike in privacy-preserving data mining, where noises are independently generated (at random) leading to imprecise aggregated results, the noises in CPDA are carefully designed to leverage the cooperation between different sensor nodes, such that the precise aggregated values can be obtained by the aggregator. The CPDA protocol classifies sensor nodes into two types: cluster leaders and cluster members. There is a one-to-many mapping between the cluster leaders and cluster members. The cluster leaders are responsible for aggregating data received from the cluster members. For security, the messages communicated between the cluster leaders and the cluster members are encrypted using different symmetric keys for each pair of nodes.

The details of the CPDA scheme are provided briefly in the following sub-sections.

### 2.1 The network model

The sensor network is modeled as a connected graph $G(V, E)$, where $V$ represents the set of senor nodes and $E$ represents the set of wireless links connecting the sensor nodes. The number of sensor nodes is taken as $|V| = N$.

A data aggregation function is taken that aggregates the individual sensor readings. CPDA scheme has focused on additive aggregation function: $f(t) = \sum_{i=1}^{N} d_i(t)$, where $d_i(t)$ is the individual sensor reading at time instant $t$ for node $i$. For computation of the aggregate functions, the following requirements are to be satisfied: (i) privacy of the individual sensor data is to be protected, i.e., each node's data should be known to no other nodes except the node itself, (ii) the number of messages transmitted within the WSN for the purpose of data aggregation should be kept at a minimum, and (iii) the aggregation result should be as accurate as possible.

**2.2 Key distribution and management**

CPDA uses a random key distribution mechanism proposed in (Eschenauer & Gligor, 2002) for encrypting messages to prevent message eavesdropping attacks. The key distribution scheme has three phases: (i) key pre-distribution, (ii) shared-key discovery, and (iii) path-key establishment. These phases are described briefly as follows.

A large key-pool of $K$ keys and their identities are first generated in the key pre-distribution phase. For each sensor nodes, $k$ keys out of the total $K$ keys are chosen. These $k$ keys form a *key ring* for the sensor node.

During the key-discovery phase, each sensor node identifies which of its neighbors share a common key with itself by invoking and exchanging discovery messages. If a pair of neighbor nodes share a common key, then it is possible to establish a secure link between them.

In the path-key establishment phase, an end-to-end path key is assigned to the pairs of neighboring nodes who do not share a common key but can be connected by two or more multi-hop secure links at the end of the shared-key discovery phase.

At the end of the key distribution phase, the probability that any pair of nodes possess at least one common key is given by (1).

$$p_{connect} = 1 - \frac{((K-k)!)^2}{(K-2k)!K!} \quad (1)$$

If the probability that any other node can overhear the encrypted message by a given key is denoted as $p_{overhear}$, then $p_{overhear}$ is given by (2).

$$p_{overhear} = \frac{k}{K} \quad (2)$$

It has been shown in (He et al., 2007) that the above key distribution algorithm is efficient for communication in a large-scale sensor network and when a limited number of keys are available for encryption of the messages to prevent eavesdropping attacks.

**2.3 Cluster-based private data aggregation (CPDA) protocol**

The CPDA scheme works in three phases: (i) cluster formation, (ii) computation of aggregate results in clusters, and (ii) cluster data aggregation. These phases are described below.

Cluster formation: Fig. 1 depicts the cluster formation process. A query server $Q$ triggers a query by sending a *HELLO* message. When the *HELLO* message reaches a sensor node, it elects itself as a cluster leader with a pre-defined probability $p$. If the value of $p$ is large, there will be more number of nodes which will elect themselves as cluster leaders. This will result in higher number of clusters in the network. On the other hand, smaller values of $p$ will lead to less number of clusters due to fewer number of cluster leader nodes. Hence, the value of the parameter $p$ can be suitably chosen to control the number of clusters in the network. If a node becomes a cluster leader, it forwards the *HELLO* message to its neighbors; otherwise, it waits for a threshold period of time to check whether any *HELLO* message arrives at it from any of its neighbors. If any *HELLO* message arrives at the node, it decides to join the cluster formed by its neighbor by broadcasting a *JOIN* message as shown in Fig. 2. This process is repeated and multiple clusters are formed so that the entire WSN becomes a collection of a set of clusters.

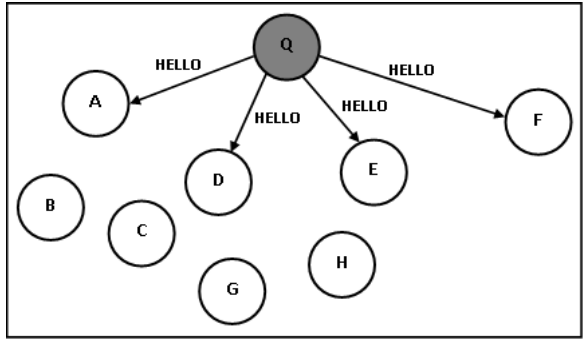

Fig. 1. The query server $Q$ sends *HELLO* messages for initiating the cluster formation procedure to its neighbors $A$, $D$, $E$ and $F$. The query server is shaded in the figure.

**Computation within clusters:** In this phase, aggregation is done in each cluster. The computation is illustrated with the example of a simple case where a cluster contains three members: $A$, $B$, and $C$, where $A$ is the assumed to be the cluster leader and the aggregator node, whereas $B$ and $C$ are the cluster member nodes. Let $a$, $b$, $c$ represent the private data held by the nodes $A$, $B$, and $C$ respectively. The goal of the aggregation scheme is to compute the sum of $a$, $b$ and $c$ without revealing the private values of the nodes.

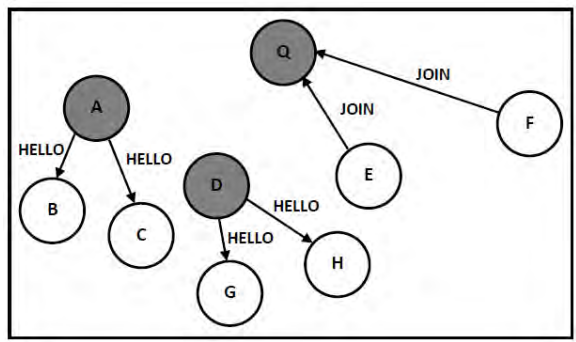

Fig. 2. $A$ and $D$ elect themselves as the cluster leaders randomly and in turn send *HELLO* messages to their neighbors. $E$ and $F$ join the cluster formed by $Q$. $B$ and $C$ join the cluster formed with $A$ as the cluster leader, while $G$ and $H$ join the cluster with $D$ as the cluster leader. All the cluster leaders and the query server are leader.

As shown in Fig. 3, for the privacy-preserving additive aggregation function, the nodes A, B, and C are assumed to share three public non-zero distinct numbers, which are denoted as $x$, $y$, and $z$ respectively. In addition, node A generates two random numbers $r_1^A$ and $r_2^A$, which are known only to node A. Similarly, nodes B and C generate $r_1^B$, $r_2^B$ and $r_1^C$, $r_2^C$ respectively, which are private values of the nodes which have generated them.

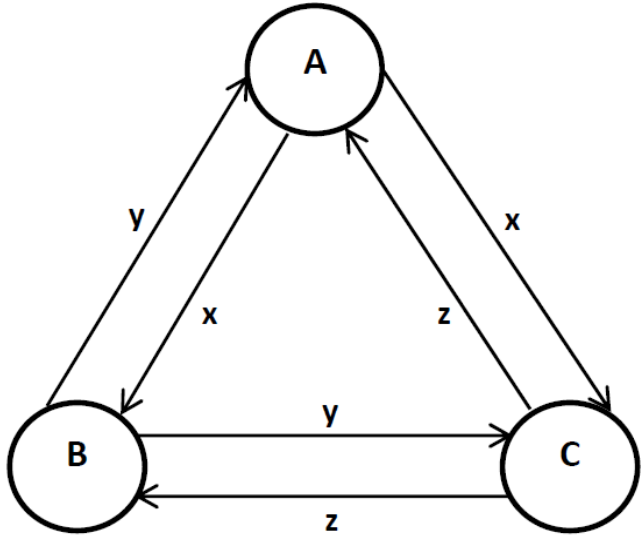

Fig. 3. Nodes A, B and C broadcast their distinct and non-zero public seeds $x$, $y$ and $z$ respectively

Node A computes $v_A^A$, $v_B^A$, and $v_C^A$ as shown in (3).

$$\left. \begin{array}{l} v_A^A = a + r_1^A x + r_2^A x^2 \\ v_B^A = a + r_1^A y + r_2^A y^2 \\ v_C^A = a + r_1^A z + r_2^A z^2 \end{array} \right\} \quad (3)$$

Similarly, node B computes $v_A^B$, $v_B^B$, and $v_C^B$ as in (4).

$$\left. \begin{array}{l} v_A^B = b + r_1^B x + r_2^B x^2 \\ v_B^B = b + r_1^B y + r_2^B y^2 \\ v_C^B = b + r_1^B z + r_2^B z^2 \end{array} \right\} \quad (4)$$

Likewise, node C computes $v_A^C$, $v_B^C$, and $v_C^C$ as in (5).

$$\left. \begin{array}{l} v_A^C = c + r_1^C x + r_2^C x^2 \\ v_B^C = c + r_1^C y + r_2^C y^2 \\ v_C^C = c + r_1^C z + r_2^C z^2 \end{array} \right\} \quad (5)$$

Node A encrypts $v_B^A$ and sends it to node B using the shared key between node A and node B. Node A also encrypts $v_C^A$ and sends it to node C using the shared key between node A and node C. In the same manner, node B sends encrypted $v_A^B$ to node A and $v_C^B$ to node C; node C sends encrypted $v_A^C$ and $v_B^C$ to node A and node B respectively. The exchanges of these encrypted messages are depicted in Fig. 4. On receiving $v_A^B$ and $v_A^C$, node A computes the sum of $v_A^A$ (already computed by node A), $v_A^B$ and $v_A^C$. Now, node A computes $F_A$ using (6).

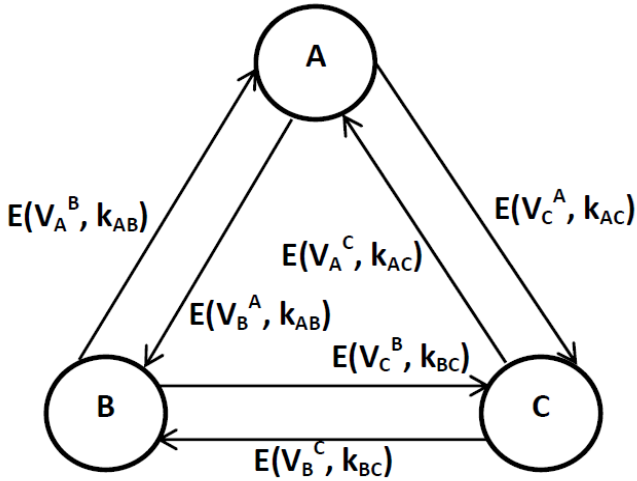

Fig. 4. Exchanges of encrypted messages among nodes A, B and C using shared keys

$$F_A = v_A^A + v_A^B + v_A^C = (a+b+c) + r_1 x + r_2 x^2 \qquad (6)$$

In (6), $r_1 = r_1^A + r_1^B + r_1^C$ and $r_2 = r_2^A + r_2^B + r_2^C$. Similarly, node B and node C compute $F_B$ and $F_C$ respectively, where $F_B$ and $F_C$ are given by (7) and (8) respectively.

$$F_B = v_B^A + v_B^B + v_B^C = (a+b+c) + r_1 y + r_2 y^2 \qquad (7)$$

$$F_C = v_C^A + v_C^B + v_C^C = (a+b+c) + r_1 z + r_2 z^2 \qquad (8)$$

Node B and node C broadcast $F_B$ and $F_C$ to the cluster leader node A, so that node A has the knowledge of the values of $F_A$, $F_B$ and $F_C$. From these values the cluster leader node A can compute the aggregated value $(a + b + c)$ as explained below.

The equations (6), (7), and (8) can be rewritten as in (9).

$$U = G^{-1} F \qquad (9)$$

In (9), $G = \begin{bmatrix} 1 & x & x^2 \\ 1 & y & y^2 \\ 1 & z & z^2 \end{bmatrix}$, $U = \begin{bmatrix} a+b+c \\ r_1 \\ r_2 \end{bmatrix}$ and $F = \begin{bmatrix} F_A & F_B & F_C \end{bmatrix}^T$.

Since $x$, $y$, $z$, $F_A$, $F_B$, and $F_C$ are known to the cluster leader node $A$, it can compute the value of $(a + b + c)$ without having any knowledge of $b$ and $c$.

In order to avoid eavesdropping attack by neighbor nodes, it is necessary to encrypt the values of $v_B^A$, $v_C^A$, $v_A^B$, $v_C^B$, $v_A^C$, and $v_B^C$. If node $B$ overhears the value of $v_C^A$, then node $B$ gets access to the values of $v_C^A$, $v_B^A$ and $F_A$. Then node $B$ can deduce: $v_A^A = F_A - v_B^A - v_C^A$. Having the knowledge of $v_A^A$, node $B$ can further obtain the value of $a$ if $x$, $v_A^A$, $v_A^B$ and $v_A^C$ are known. However, if node $A$ encrypts $v_C^A$ and sends it to node $C$, then node $B$ cannot get $v_C^A$. With the knowledge of $v_B^A$, $F_A$, and $x$ from node $A$, node $B$ cannot deduce the value of $a$. If node $B$ and node $C$ collude and reveal node $A$'s information (i.e., $v_B^A$ and $v_C^A$), to each other, then node $A$'s privacy will be compromised and its private value $a$ will be revealed. In order to reduce the probability of such collusion attacks, the cluster size should be as large as possible, since in a cluster of size $m$, at least $(m - 1)$ nodes should collude in order to successfully launch the attack. Higher values of $m$ will require larger number of colluding nodes thereby making the attack more difficult.

Cluster data aggregation The CPDA scheme has been implemented on top of a protocol known as *Tiny Aggregation* (TAG) protocol (Madden et al., 2002). Using the TAG protocol, each cluster leader node routes the sum of the values in the nodes in its cluster to the query server through a TAG routing tree whose root is situated at the server.

## 3. An Attack on the CPDA scheme

In this section, we present an efficient attack (Sen & Maitra, 2011) on the CPDA aggregation scheme. The objective of the attack is to show the vulnerability of the CPDA scheme which can be suitably exploited by a malicious participating sensor node. The intention of the malicious node is to participate in the scheme in such a way that it can get access to the private values (i.e., $a$, $b$ and $c$) of the participating sensor nodes. For describing the attack scenario, we use the same example cluster consisting of three sensor nodes $A$, $B$ and $C$. Node $A$ is the cluster leader whereas node $B$ and node $C$ are the cluster members. We distinguish two types of attacks: (i) attack by a malicious cluster leader (e.g., node $A$) and (ii) attack by a malicious cluster member (e.g., either node $B$ or node $C$). These two cases are described in detail in the following sub-sections.

### 3.1 Privacy attack by a malicious cluster leader node

Let us assume that the cluster leader node $A$ is malicious. Node $A$ chooses a very large value of $x$ such that $x \gg y, z$. Since $y$ and $z$ are public values chosen by node $B$ and node $C$ which are broadcast in the network by node $B$ and node $C$ respectively, it is easy for node $A$ to choose a suitable value for $x$.

Nodes $A$, $B$ and $C$ compute the values of $v_A^A$, $v_B^A$, $v_C^A$, $v_A^B$, $v_B^B$, $v_C^B$, $v_A^C$, $v_B^C$, and $v_C^C$ using (3), (4) and (5) as described in Section 2.3. As per the CPDA scheme, node $A$ receives:

$v_A^B = b + r_1^B x + r_2^B x^2$ from node B. Since $x$ is very large compared to $b$ and $r_1^B$ node A can derive the value of $r_2^B$ using (10) where we consider integer division.

$$\frac{v_A^B}{x^2} = \frac{b}{x^2} + \frac{r_1^B}{x} + r_2^B = 0 + 0 + r_2^B = r_2^B \qquad (10)$$

Using the value of $r_2^B$ as derived in (10), and using $v_A^B = b + r_1^B x + r_2^B x^2$, node A can now compute the value of $r_1^B$ by solving (11).

$$\frac{v_A^B - r_2^B x^2}{x} = \frac{b}{x} + r_1^B = 0 + r_1^B = r_1^B \qquad (11)$$

In the same manner, node A derives the values of $r_1^C$ and $r_2^C$ from $v_A^C$ received from node C. Since $r_1 = r_1^A + r_1^B + r_1^C$, and $r_2 = r_2^A + r_2^B + r_2^C$, as shown in (6), (7) and (8), node A can compute the values of $r_1$ and $r_2$ ($r_1^B$, $r_2^B$, $r_1^C$, and $r_2^C$ are derived as shown above, and $r_1^A$ and $r_2^A$ were generated by node A).

At this stage, node A uses the values of $F_B$ and $F_C$ received from node B and node C respectively as shown in (7) and (8). Node A has now two linear simultaneous equations with two unknowns: $b$ and $c$, the values of $y$ and $z$ being public. Solving (7) and (8) for $b$ and $c$, the malicious cluster leader node A can get the access to the private information.

### 3.2 Privacy attack by a malicious cluster member node

In this scenario, let us assume that the cluster member node B is malicious and it tries to access the private values of the cluster leader node A and the cluster member node C. Node B chooses a very large value of $y$ so that $y >> x, z$. Once the value of $F_B$ is computed in (7), node B derives the value of $r_2$ and $r_1$ using (12) and (13).

$$\frac{F_B}{y^2} = \frac{(a+b+c)}{y^2} + \frac{r_1}{y} + r_2 = 0 + 0 + r_2 \qquad (12)$$

$$\frac{F_B - r_2 y^2}{y} = \frac{(a+b+c)}{y} + r_1 = 0 + r_1 = r_1 \qquad (13)$$

As per the CPDA scheme, node B receives $v_B^C = c + r_1^C y + r_2^C y^2$ from node C. Since the magnitude of $y$ is very large compared to $c$, $r_1^C$ and $r_2^C$, it is easy for node B to derive the values of $r_2^C$ and $r_1^C$ using (14) and (15) respectively.

$$\frac{v_B^C}{y^2} = \frac{c}{y^2} + \frac{r_1^C}{y} + r_2^C = 0 + 0 + r_2^C = r_2^C \qquad (14)$$

$$\frac{v_B^C - r_2^C y^2}{y} = \frac{c}{y} + r_1^C = 0 + r_1^C = r_1^C \qquad (15)$$

Using (12), (13), (14}) and (15) node B can compute $r_1^A = r_1 + r_1^B - r_1^C$ and $r_2^A = r_2 - r_2^B - r_2^C$. Now, node B can compute the value of $a$ using $v_A^B = a + r_1^A y + r_2^A y^2$ (received from node A), in which the values of all the variables are known except that of *a*. In a similar fashion, node *B* derives the value of *c* using $v_B^C = c + r_1^C y + r_2^C y^2$ (received from node *C*).

Since the private values of the nodes *A* and *C* are now known to node *B*, the privacy attack launched by participating cluster member node *B* is successful on the CPDA aggregation scheme.

## 4. Modification of the CPDA Scheme

In this section, we present two modifications of CPDA scheme: one towards making the protocol more efficient and the other for making it more secure.

### 4.1 Modification of CPDA scheme for enhanced efficiency

In this section, a modification is proposed for the CPDA protocol for achieving enhanced efficiency in its operation. The modification is based on suitable choice for the value of *x* (the public seed) done by the aggregator node *A*.

Let us assume that the node *A* chooses a large value of *x* such that the following conditions in (16) and (17) are satisfied.

$$r_2 x^2 \gg r_1 x \tag{16}$$

$$r_1 x \gg (a + b + c) \tag{17}$$

In (16) and (17), $r_1 = r_1^A + r_1^B + r_1^C$ and $r_2 = r_2^A + r_2^B + r_2^C$. Now, node *A* has computed the value of $F_A$ as shown in (6). In order to efficiently compute the value of (*a* + *b* + *c*), node *A* divides the value of $F_A$ by $x^2$ as shown in (18).

$$\frac{F_A}{x^2} = \frac{(a+b+c)}{x^2} + \frac{r_1 x}{x^2} + r_2 = 0 + 0 + r_2 = r_2 \tag{18}$$

Using (18), node *A* derives the value of $r_2$. Once the value of $r_2$ is deduced, node *A* attempts to compute the value of $r_1$ using (19) and (20).

$$F_A - r_2 x^2 = (a+b+c) + r_1 x \tag{19}$$

$$r_1 = \frac{(F_A - r_2 x^2)}{x} - \frac{(a+b+c)}{x} = \frac{(F_A - r_2 x^2)}{x} - 0 = \frac{(F_A - r_2 x^2)}{x} \tag{20}$$

Since, the values of $F_A$, $r_2$ and *x* are all known to node *A*, it can compute the value of $r_1$ using (20). Once the values of $r_1$ and $r_2$ are computed by node *A*, it can compute the value of (*a* + *b* + *c*) using (6). Since the computation of the sum (*a* + *b* + *c*) by node *A* involves two division operations (involving integers) only (as done in (18) and (20)), the modified CPDA scheme is light-weight and it is much more energy-efficient hence much more energy- and time-efficient as compared to the original CPDA scheme. The original CPDA scheme involved additional computations of the values of $F_B$ and $F_C$, and an expensive matrix inversion operation as described in Section 2.3.

### 4.2 Modification of the CPDA scheme for resisting the attack

In this section, we discuss the modifications required on the existing CPDA scheme so that a malicious participant node cannot launch the attack described in Section 3.

It may be noted that, the vulnerability of the CPDA scheme lies essentially in the unrestricted freedom delegated on the participating nodes for generating their public seed values. For example, nodes *A*, *B* and *C* have no restrictions on their choice for values of $x$, $y$ and $z$ respectively while they generate these values. A malicious attacker can exploit this freedom to generate an arbitrarily large public seed value, and can thereby launch an attack as discussed in Section 3.

In order to prevent such an attack, the CPDA protocol needs to be modified. In this modified version, the nodes in a cluster make a check on the generated public seed values so that it is not possible for a malicious participant to generate any arbitrarily large seed value. For a cluster with three nodes, such a constraint may be imposed by the requirement that the sum of any two public seeds must be greater than the third seed. In other words: $x + y > z$, $z + x > y$, and $y + z > x$. If these constraints are satisfied by the generated values of $x$, $y$ and $z$, it will be impossible for any node to launch the attack and get access to the private values of the other participating nodes.

However, even if the above restrictions on the values of $x$, $y$ and $z$ are imposed, the nodes should be careful in choosing the values for their secret random number pairs. If two nodes happen to choose very large values for their random numbers compared to those chosen by the third node, then it will be possible for the third node to get access to the private values of the other two nodes. For example, let us assume that nodes *A* and *C* have chosen the values of $r_1^A$, $r_2^A$ and $r_1^C$, $r_2^C$ such that they are all much larger than $r_1^B$ and $r_2^B$ - the private random number pair chosen by node *B*. It will be possible for node *B* to derive the values of *a* and *c*: the private values of nodes *A* and *C* respectively. This is explained in the following.

Node *B* receives $v_B^A = a + r_1^A y + r_2^A y^2$ from node *A* and computes the values of $r_1^A$ and $r_2^A$ using (21) and (22).

$$\frac{v_B^A}{y^2} = \frac{a}{y^2} + \frac{r_1^A}{y} + r_2^A = 0 + 0 + r_2^A \tag{21}$$

$$\frac{v_B^A - r_2^A y^2}{y} = \frac{a}{y} + r_1^A = 0 + r_1^A = r_1^A \tag{22}$$

In a similar fashion, node *B* derives the values of $r_1^C$ and $r_2^C$ from $v_B^C$ received from node *C*. Now, node *B* computes $r_1 = r_1^A + r_1^B + r_1^C$ and $r_2 = r_2^A + r_2^B + r_2^C$, since it has access to the values of all these variables. In the original CPDA scheme in (He et al., 2007), the values of $F_B$ and $F_C$ are broadcast by nodes *B* and *C* in unencrypted from. Hence, node *B* has access to both these values. Using (7) and (8), node *B* can compute the values of *a* and *c*, since these are the only unknown variables in the two linear simultaneously equations.

In order to defend against the above vulnerability, the CPDA protocol needs further modification. In this modified version, after the values $v_A^A$, $v_A^B$, and $v_A^C$ are generated and shared by nodes *A*, *B* and *C* respectively, the nodes check whether the following constraints are satisfied: $v_A^A + v_A^B > v_A^C$, $v_A^B + v_A^C > v_A^A$, and $v_A^C + v_A^A > v_A^B$. The nodes proceed for further execution of the algorithm only if the above three inequalities are satisfied. If all three inequalities are not satisfied, there will be a possibility that the random numbers generated by one node is much larger than those generated by other nodes - a scenario which indicates a possible attack by a malicious node.

## 5. Performance analysis

In this section, we present a brief comparative analysis of the overheads of the original CPDA protocol and the proposed modified CPDA protocols that we have discussed in Section 4.1 and Section 4.2. Our analysis is based on two categories of overheads: (i) overhead due to message communication in the network and (ii) computational overhead at the sensor nodes.

### 5.1 Communication overhead

We compare communication overheads of three protocols - the *tiny aggregation protocol* (TAG), the original CPDA protocol and the proposed modified CPDA protocols. In TAG, each sensor node needs to send 2 messages for the data aggregation protocol to work. One *HELLO* message communication from each sensor node is required for forming the aggregation tree, and one message is needed for data aggregation. However, this protocol only performs data aggregation and does not ensure any privacy for the sensor data. In the original CPDA protocol, each cluster leader node sends 4 messages and each cluster member node sends 3 messages for ensuring that the aggregation protocol works in a privacy-preserving manner. In the example cluster shown in Fig. 3, the 4 messages sent by the cluster leader node *A* are: one *HELLO* message for forming the cluster, one message for communicating the public seed *x*, one message for communicating $v_B^A$ and $v_C^A$ to cluster member nodes *B* and *C* respectively, and one message for sending the aggregate result from the cluster. Similarly, the 3 messages sent by the cluster member node *B* are: one message for communicating its public seed *y*, one message for communicating $v_A^B$ and $v_C^B$ to cluster leader node *A* and cluster member node *C* respectively, and one message for communicating the intermediate result $F_B$ to the cluster leader node *A*.

In contrast to the original CPDA protocol, the modified CPDA protocol in Section 4.1 involves 3 message communications from the cluster leader node and 2 message communications from each cluster member node. The 3 messages sent by the cluster leader node *A* are: one *HELLO* message for forming the cluster, one message for broadcasting its public seed *x*, and one message for sending the final aggregate result. It may be noted that in this protocol, the cluster leader node *A* need not send $v_B^A$ and $v_C^A$ to the cluster member nodes *B* and *C* respectively. Each cluster member node needs to send 2 messages. For example, the cluster member node *B* needs to broadcast its public seed *y*, and also needs to send $v_A^B$ to the cluster leader node *A*. Unlike in the original CPDA protocol, the cluster member node *B* does not send $F_B$ to the cluster leader. Similarly, the cluster member node *C* does not send $F_C$ to the cluster leader node *A*. In a cluster consisting of three members, the original CPDA protocol would involve 10 messages (4 messages from the cluster leader and 3 messages from each cluster member). The modified CPDA protocol presented in Section 4.1, on the other hand, would involve 7 messages (3 messages from the cluster leader and 2 messages from each cluster member) in a cluster of three nodes. Therefore, in a cluster of three nodes, the modified CPDA protocol presented in Section 4.1 will involve 3 less message communications. Since in a large-scale WSN the number of clusters will be quite high, there will be an appreciable reduction in the communication overhead in the modified CPDA protocol presented in Section 4.1.

The secure version of the modified CPDA protocol presented in Section 4.2 involves the same communication overhead as the original CPDA protocol. However, if any node chooses abnormally higher values for its public seed or its private random numbers, the secure version of the modified CPDA protocol will involve 2 extra messages from each of the participating sensor nodes. Therefore, in a cluster of three nodes, the secure version of the modified CPDA protocol will involve 6 extra messages in the worst case scenario when compared with the original CPDA protocol.

If $p_c$ is the probability of a sensor node electing itself as a cluster leader, the average number of messages sent by a sensor node in the original CPDA protocol is: $4p_c + 3(1-p_c) = 3 + p_c$. Thus, the message overhead in the original CPDA is less than twice as that in TAG. However, in the modified CPDA protocol presented in Section 4.1, the average number of messages communicated by a sensor node is: $3p_c + 2(1-p_c) = 2 + p_c$. As mentioned in Section 2.3, in order to prevent collusion attack by sensor nodes, the cluster size in the CPDA protocol should be as large as possible. This implies that the value of $p_c$ should be small. Since the value of $p_c$ is small, it is clear that the message overhead in the modified CPDA protocol presented in Section 4.1 is almost the same as that in TAG and it is much less (one message less for each sensor node) than that of the original CPDA protocol. In the secure version of the protocol in Section 4.2, the communication overhead, in the average case, will be the same as in the original CPDA protocol. However, in the worst case, the number of messages sent by a sensor node in this protocol will be: $6p_c + 5(1-p_c) = 5 + p_c$. This is 2.5 times the average communication overhead in the TAG protocol and 1.67 times the average communication overhead in the original CPDA protocol. The secure protocol, therefore, will involve 67% more overhead in the worst case scenario (where a malicious participant sensor node chooses abnormally higher values for its public seed as well as for its private random numbers).

### 5.2 Computational overhead

In this section, we present a comparative analysis of the computational overheads incurred by the sensor nodes in the original CPDA protocol and in the proposed efficient version of the protocol.

**Computational overhead of the original CPDA protocol:** The computational overhead of the CPDA protocol can be broadly classified into four categories: (i) computation of the parameters, (ii) computation for encrypting messages, (iii) computation of the intermediate results, and (iv) computation of the final aggregate result at the cluster leader node. The details of these computations are presented below:

i.  *Computation of the parameters at the sensor nodes:* Each sensor node in a three member cluster computes three parameters. For example, the cluster leader node *A* computes $v_A^A$, $v_B^A$, $v_C^A$. Similarly, the cluster member node *B* computes $v_A^B$, $v_B^B$ and $v_C^B$. We first compute the overhead due these computations.

Since $v_A^A = a + r_1^A x + r_2^A x^2$, for computation of $v_A^A$, node *A* needs to perform 2 addition, 2 multiplication and 1 exponentiation operations. Hence, for computing $v_A^A$, $v_B^A$ and $v_C^A$, node *A* needs to perform 6 addition, 6 multiplication and 3 exponentiation operations. Therefore, in a cluster consisting of three members, for computation of all parameters, the original CPDA protocol requires 18 addition, 18 multiplication and 9 exponentiation operations.

ii. *Computations for encrypting messages*: Some of the messages in the CPDA protocol need to be communicated in encrypted form. The encryption operation involves computational overhead. For example, node *A* needs to encrypt $v_B^A$ and $v_C^A$ before sending them to nodes *B* and *C* respectively. Therefore, 2 encryption operations are required at node *A*. For a cluster consisting of three members, the CPDA protocol will need 6 encryption operations.

iii. *Computations of intermediate results*: The nodes *A*, *B*, and *C* need to compute the intermediate values $F_A$, $F_B$ and $F_C$ respectively for computation of the final aggregated result. Since $F_A = v_A^A + v_A^B + v_A^C = (a + b + c) + r_1 x + r_2 x^2$ and $r_1 = r_1^A + r_1^B + r_1^C$ and $r_2 = r_2^A + r_2^B + r_2^C$, for computing $F_A$, node *A* will need to perform 4 addition operations. Therefore, for a cluster of three members, 12 addition operations will be needed.

iv. *Aggregate computation at the cluster leader*: For computing the final aggregated result in a privacy-preserving way, the cluster leader node *A* needs to perform one matrix inversion operation and one matrix multiplication operation.

The summary of various operations in the original CPDA protocol are presented in Table 1.

| Operation Type        | No. of operations |
|-----------------------|-------------------|
| Addition              | 30                |
| Multiplication        | 18                |
| Exponentiation        | 3                 |
| Encryption            | 6                 |
| Matrix multiplication | 1                 |
| Matrix inversion      | 1                 |

Table 1. Operation in the CPDA protocol

**Computational overhead of the modified CPDA protocol:** The overhead of the efficient version of the CPDA protocol presented in Section 4.1 are due to: (i) computation of the parameters at the sensor nodes, (ii) computation of the intermediate result at the cluster leader node, and (iii) computation of the aggregated result at the cluster leader node. The details of these computations are presented below.

i. *Computation of the parameters at the sensor nodes*: In the modified version of the CPDA protocol, the nodes *A*, *B* and *C* need to only compute $v_A^A$, $v_A^B$, and $v_A^C$ respectively. As shown earlier, each parameter computation involves 2 addition, 2 multiplication and 1 exponentiation operations. Therefore, in total, 6 addition, 6 multiplication, and 3 exponentiation operations will be needed.

ii. *Computations for encrypting messages*: The nodes *B* and *C* will need to encrypt the messages $v_A{}^B$ and $v_A{}^C$ respectively before sending them to the cluster leader node *A*. Therefore, 2 encryption operations will be required.
iii. *Computation of intermediate result*: The cluster leader node *A* will only compute $F_A$ in the modified CPDA. The cluster member nodes *B* and *C* need not perform any computations here. As discussed earlier, computation of $F_A$ needs 4 addition operations.
iv. *Aggregate computation at the cluster leader*: For computation of the final result at the cluster leader node, 2 integer division and 2 subtraction operations will be required.
v. The summary of various operations in the modified CPDA protocol are presented in Table 2.

| Operation Type | No. of operations |
|---|---|
| Addition | 10 |
| Subtraction | 2 |
| Multiplication | 6 |
| Division | 2 |
| Exponentiation | 3 |
| Encryption | 2 |

Table 2. Operation in the proposed modified CPDA protocol

It is clearly evident from Table 1 and Table 2 that the modified version of the CPDA protocol involves much less computational overhead than the original version of the protocol.

## 6. Security requirements in data aggregation protocols for WSNs

The purpose of any WSN deployment is to provide the users with access to the information of interest from the data gathered by spatially distributed sensor nodes. In most applications, users require only certain aggregate functions of this distributed data. Examples include the average temperature in a network of temperature sensors, a particular trigger in the case of an alarm network, or the location of an event. Such aggregate functions could be computed under the end-to-end information flow paradigm by communicating all relevant data to a central collector node. This, however, is a highly inefficient solution for WSNs which have severe constraints in energy, memory and bandwidth, and where tight latency constraints are to be met. As mentioned in Section 1 of this chapter, an alternative solution is to perform in-network computations (Madden et al., 2005). However, in this case, the question that arises is how best to perform the distributed computations over a network of nodes with wireless links. What is the optimal way to compute, for example, the average, min, or max of a set of statistically correlated values stored in different nodes? How would such computations be performed in the presence of unreliability such as noise, packet drops, and node failures? Such questions combine the complexities of multi-terminal information theory, distributed source coding, communication complexity, and distributed computation. This makes development of an efficient in-network computing framework for WSNs very challenging.

Apart from making a trade-off between the level of accuracy in aggregation and the energy expended in computation of the aggregation function, another issue that needs serious attention in WSN is security. Unfortunately, even though security has been identified as a major challenge for sensor networks (Karlof & Wagner, 2003), most of the existing proposals for data aggregation in WSNs have not been designed with security in mind. Consequently, these schemes are all vulnerable to various types of attacks (Sen, 2009). Even when a single sensor node is captured, compromised or spoofed, an attacker can often manipulate the value of an aggregate function without any bound, gaining complete control over the computed aggregate. In fact, any protocol that computes the average, sum, minimum, or maximum function is insecure against malicious data, no matter how these functions are computed. To defend against these critical threats, in this chapter, an energy-efficient aggregation algorithm based on distributed estimation approach. The algorithm is secure and robust against malicious attacks in WSNs. The main threat that has been considered while designing the proposed scheme is the *injection of malicious data* in the network by an adversary who has compromised a sensor's sensed value by subjecting it to unusual temperature, lighting, or other spoofed environmental conditions. In designing the proposed algorithm, a WSN is considered as a collective entity that performs a sensing task and have proposed a distributed estimation algorithm that can be applied to a large class of aggregation problems.

In the proposed scheme (Sen, 2011), each node in a WSN has complete information about the parameter being sensed. This is in contrast to the snapshot aggregation, where the sensed parameters are aggregated at the intermediate nodes till the final aggregated result reaches the root. Each node, in the proposed algorithm, instead of unicasting its sensed information to its parent, broadcasts its estimate to all its neighbors. This makes the protocol more fault-tolerant and increases the information availability in the network. The scheme is an extension of the one suggested in (Boulis et al., 2003). However, it is more secure and reliable even in presence of compromised and faulty nodes in a WSN.

In the following section, we provide a brief discussion on some of the well-known secure aggregation schemes for WSNs.

## 7. Overview of some aggregation protocols for WSNs

Extensive work has been done on aggregation applications in WSNs. However, security and energy- two major aspects for design of an efficient and robust aggregation algorithm have not attracted adequate attention. Before discussing some of the existing secure aggregation mechanisms, we present a few well-known aggregation schemes for WSNs.

In (Heidemann, 2001), a framework for flexible aggregation in WSNs has been presented following snapshot aggregation approach without addressing issues like energy efficiency and security in the data aggregation process. A cluster-based algorithm has been proposed in (Estrin et al., 1999) that uses *directed diffusion* technique to gather a global perspective utilizing only the local nodes in each cluster. The nodes are assigned different level – level 0 being assigned to the nodes lying at the lowest level. The nodes at the higher levels can communicate with the nodes in the same cluster and the cluster head node. This effectively enables *localized cluster computation*. The nodes at the higher level communicate the local information of the cluster to get a global picture of the network aggregation. In (Madden et al., 2002), the authors have proposed a mechanism called TAG – a generic data aggregation scheme that involves a language similar to SQL for generating queries in a WSN. In this scheme, the *base station* (BS) generates a query using the query language, and the sensor nodes send their reply using routes constructed based on a *routing tree*. At each point in the routing tree, the data is aggregated using some aggregation function that was defined in the initial query sent by the BS. In (Shrivastava et al., 2004), a summary structure for supporting fairly complex aggregate functions, such as median and range quires have been proposed. Computation of relatively easier function such as min/max, sum, and average are also supported in the proposed framework. However, more complex aggregates, such as the most frequently reported data values are not supported. The computed aggregate functions are approximate but the estimate errors are statistically bounded. There are also propositions based on programmable sensor networks for aggregation based on snapshot algorithms (Jaikaeo et al., 2000). In (Zhao et al., 2002), the authors have focussed their attention into the problem of providing a residual energy map of a WSN. They have proposed a scheme for computing the equi-potential curves of residual energy with certain acceptable margin of error. A simple but efficient aggregation function is proposed where the location approximation of the nodes are not computed. A more advanced aggregate function can be developed for this purpose that will encompass an accurate convex curve. For periodic update of the residual energy map, the authors have proposed a naïve scheme of incremental updates. Thus if a node changes its value beyond the tolerance limit its value is transmitted and aggregated again by some nodes before the final change reaches the user. No mechanism exists for prediction of changes or for estimation of correlation between sensed values for the purpose of setting the tolerance threshold. In (Goel & Imielinski, 2001), a scheme has been proposed for the purpose of monitoring the sensed values of each individual sensor node in a WSN. There is no aggregation algorithm in the scheme; however, the spatial-temporal correlation between the sensed data can be extrapolated to fit an aggregation function. The authors have also attempted to modify the techniques of MPEG-2 for sensor network monitoring to optimize communication overhead and energy. A central node computes predictions and transmits them to all the nodes. The nodes send their update only if their sensed data deviate significantly from the predictions. A distributed computing framework is developed by establishing a hierarchical dependency among the nodes. An energy efficient aggregation algorithm is proposed by the authors in (Boulis et al., 2003), in which each node in a WSN senses the parameter and there is no hierarchical dependency among the nodes. The nodes in a neighbourhood periodically broadcast their information based on a threshold value.

As mentioned earlier in this section, none of the above schemes consider security aspects in the aggregation schemes. Security in aggregation schemes for WSNs has also attracted attention from the researchers and a considerable number of propositions exist in the literature in this perspective. We discuss some of the well-known mechanisms below.

A *secure aggregation* (SA) protocol has been proposed that uses the *μTESLA* protocol (Hu & Evans, 2003). The protocol is resilient to both intruder devices and single device key compromises. In the proposition, the sensor nodes are organized into a tree where the internal nodes act as the aggregators. However, the protocol is vulnerable if a parent and one of its child nodes are compromised, since due to the delayed disclosure of symmetric keys, the parent node will not be able to immediately verify the authenticity of the data sent by its children nodes.

Przydatek et al. have presented a *secure information aggregation* (SIA) framework for sensor networks (Przydatek et al., 2003; Chan et al., 2007). The framework consists of three categories of node: a home server, base station and sensor nodes. A base station is a resource-enhanced node which is used as an intermediary between the home server and the sensor nodes, and it is also the candidate to perform the aggregation task. SIA assumes that each sensor has a unique identifier and shares a separate secret cryptographic key with both the home server and the aggregator. The keys enable message authentication and encryption if data confidentiality is required. Moreover, it further assumes that the home server and the base station can use a mechanism, such as *μTESLA*, to broadcast authenticated messages. The proposed solution follows *aggregate-commit-prove* approach. In the *first phase*: *aggregate*- the aggregator collects data from sensors and locally computes the aggregation result using some specific aggregate function. Each sensor shares a key with the aggregator. This allows the aggregator to verify whether the sensor reading is authentic. However, there is a possibility that a sensor may have been compromised and an adversary has captured the key. In the proposed scheme there is no mechanism to detect such an event. In the *second phase*: *commit*- the aggregator commits to the collected data. This phase ensures that the aggregator actually uses the data collected from the sensors, and the statement to be verified by the home server about the correctness of computed results is meaningful. One efficient mechanism for committing is a *Merkle hash-tree* construction (Merkle, 1980). In this method, the data collected from the sensors is placed at the leaves of a tree. The aggregator then computes a binary hash tree staring with the leaf nodes. Each internal node in the hash tree is computed as the hash value of the concatenation of its two children nodes. The root of the tree is called the commitment of the collected data. As the hash function in use is collision free, once the aggregator commits to the collected values, it cannot change any of the collected values. In the *third and final phase*, the aggregator and the home server engage in a protocol in which the aggregator communicates the aggregation result. In addition, aggregator uses an interactive proof protocol to prove correctness of the reported results. This is done in two logical steps. In the first step, the home server ensures that the committed data is a good representation of the sensor data readings collected. In the second step, the home server checks the reliability of the aggregator output. This is done by checking whether the aggregation result is close to the committed results. The interactive proof protocol varies depending on the aggregation function is being used. Moreover, the authors also presented efficient protocols for secure computation of the median and the average of the measurements, for the estimation of the network size, and for finding the minimum and maximum sensor reading.

In (Mahimkar & Rappaport, 2004), a protocol is proposed that uses elliptic curve cryptography for encrypting the data in WSNs. The scheme is based on clustering where all nodes within a cluster share a secret cluster key. Each sensor node in a cluster generates a partial signature over its data. Each aggregator aggregates its cluster data and broadcasts the aggregated data in its cluster. Each node in a cluster checks its data with the aggregated data broadcast by the aggregator. A sensor node puts its partial signature to authenticate a message only if the difference between its data and aggregated data is less than a threshold. Finally, the aggregator combines all the partially signed message s to form a full signature with the authenticated result.

Deng et al. proposed a collection of mechanisms for *securing in-network processing* (SINP) for WSNs (Deng et al., 2003). Security mechanisms have been proposed to address the downstream requirement that sensor nodes authenticate commands disseminated from parent aggregators and the upstream requirement that aggregators authenticate data produced by sensors before aggregating that data. In the downstream stage, two techniques are involved: one way functions and *μTESLA*. The upstream stage requires that a pair-wise key be shared between an aggregator and its sensor nodes.

Cam et al. proposed an *energy-efficient secure pattern-based data aggregation* (ESPDA) protocol for wireless sensor networks (Cam et al., 2003; Cam et al., 2005; Cam et al., 2006a). ESPDA is applicable for hierarchy-based sensor networks. In ESPDA, a cluster-head first requests sensor nodes to send the corresponding pattern code for the sensed data. If multiple sensor nodes send the same pattern code to the cluster-head, only one of them is permitted to send the data to the cluster-head. ESPDA is secure because it does not require encrypted data to be decrypted by cluster-heads to perform data aggregation.

Cam et al. have introduced another *secure differential data aggregation* (SDDA) scheme based on pattern codes (Cam et al., 2006b). SDDA prevents redundant data transmission from sensor nodes by implementing the following schemes: (1) SDDA transmits differential data rather than raw data, (2) SDDA performs data aggregation on pattern codes representing the main characteristics of the sensed data, and (3) SDDA employs a sleep protocol to coordinate the activation of sensing units in such a way that only one of the sensor nodes capable of sensing the data is activated at a given time. In the SDDA data transmission scheme, the raw data from the sensor nodes is compared with the reference data and the difference of them is transmitted in the network. The reference data is obtained by taking the average of previously transmitted data.

In (Sanli et al., 2004 ), a *secure reference-based data aggregation* (SRDA) protocol is proposed for cluster-based WSNs, in which raw data sensed by sensor nodes are compared with reference data values and then only difference data is transmitted to conserve sensor energy. Reference data is taken as the average of a number of historical (i.e. past) sensor readings. However, a serious drawback of the scheme is that does not allow aggregation at the intermediate nodes.

To defend against attacks by malicious aggregator nodes in WSNs which may falsely manipulate the data during the aggregation process, a cryptographic mechanism has been proposed in (Wu et al., 2007). In the proposed mechanism, a *secure aggregation tree* (SAT), is constructed that enables monitoring of the aggregator nodes. The child nodes of the aggregators can monitor the incoming data to the aggregators and can invoke a voting scheme in case any suspicious activities by the aggregator nodes are observed.

A *secure hop-by-hop data aggregation protocol* (SDAP) has been proposed in (Yang et al., 2006), in which a WSN is dynamically partitioned into multiple logical sub-trees of almost equal sizes using a probabilistic approach. In this way, fewer nodes are located under a high-level sensor node, thereby reducing potential security threats on nodes at higher level. Since a compromised node at higher level in a WSN will cause more adverse effect on data aggregation than on a lower-level node, the authors argue that by reducing number of nodes at the higher level in the logical tree, aggregation process becomes more secure.

In (Ozdemir, 2007), a secure and reliable data aggregation scheme – SELDA- is proposed that makes use of the concept of web of trust. Trust and reputation based schemes have been extensively used for designing security solutions for multi-hop wireless networks like *mobile ad hoc networks* (MANETs), *wireless mesh networks* (WMNs) and WSNs (Sen, 2010b; Sen, 2010c; Sen 2010d). In this scheme, sensor nodes exchange trust values in their neighborhood to form a *web of trust* that facilitates in determining secure and reliable paths to aggregators. Observations from the sensor nodes which belong to a web of trust are given higher weights to make the aggregation process more robust.

A *data aggregation and authentication* (DAA) protocol is proposed in (Cam & Ozdemir, 2007), to integrate false data detection with data aggregation and confidentiality. In this scheme, a monitoring algorithm has been proposed for verifying the integrity of the computed aggregated result by each aggregator node.

In order to minimize false positives (a scenario where an alert is raised, however there is no attack), in a WSN, a dynamic threshold scheme is proposed in (Parkeh & Cam, 2007), which dynamically varies the threshold in accordance with false alarm rate. A data aggregation algorithm is also proposed to determine the detection probability of a target by fusing data from multiple sensor nodes.

Du et al. proposed a *witness-based data aggregation* (WDA) scheme for WSNs to assure the validation of the data fusion nodes to the base station (Du et al., 2003). To prove the validity of the fusion results, the fusion node has to provide proofs from several witnesses. A witness is one who also conducts data fusion like a data fusion node, but does not forward its result to the base station. Instead, each witness computes the MAC of the result and then provides it to the data fusion node, which must forward the proofs to the base station. This scheme can defend against attacks on data integrity in WSNs.

Wagner studied secure data aggregation in sensor networks and proposed a mathematical framework for formally evaluating their security (Wagner, 2004). The robustness of an aggregation operator against malicious data is quantified. Ye et al. propose a *statistical en-route filtering mechanism* to detect any forged data being sent from the sensor nodes to the base station of a WSN using multiple MACs along the path from the aggregator to the base station (Ye et al., 2004; Ye et al., 2005).

## 8. The proposed distributed secure aggregation protocol

In this section, we propose a distributed estimation algorithm that is secure and resistant to insider attack by compromised and faulty nodes. There are essentially two categories of aggregation functions (Boulis et al., 2003):

- Aggregation functions that are dependent on the values of a few nodes (e.g., the *max* result is based on one node).
- Aggregation functions whose values are determined by all the nodes (e.g., the average function).

However, computation of both these types of functions are adversely affected by wrong sensed result sent by even a very few number of compromised nodes. In this chapter, we consider only the first case, i.e., aggregation function that find or approximate some kind of boundaries (e.g., maxima, minima), and hence the aggregation result is determined by the values of few nodes. However, the proposed algorithm does not assume any knowledge about the underlying physical process.

## 8.1 The proposed secure aggregation algorithm

In the proposed distributed estimation algorithm, a sensor node instead of transmitting a partially aggregated result, maintains and if required, transmits an estimation of the global aggregated result. The global aggregated description in general will be a vector since it represents multi-dimensional parameters sensed by different nodes. A global estimate will thus be a probability density function of the vector that is being estimated. However, in most of the practical situations, due to lack of sufficient information, complex computational requirement or unavailability of sophisticated estimation tools, an estimate is represented as: (*estimated value*, *confidence indication*), which in computational terms can be represented as: (*average of estimated vector*, *covariance matrix of estimated vector*). For the sake of manipulability with tools of estimation theory, we have chosen to represent estimates in the form of $(A, P_{AA})$ with $A$ being the mean of the aggregated vector and $P_{AA}$ being the covariance matrix of vector $A$. For the *max* aggregation function, vector $A$ becomes a scalar denoting the mean of the estimated max, and $P_{AA}$ becomes simply the variance of $A$.

In the snapshot aggregation, a node does not have any control on the rate at which it send information to its parents; it has to always follow the rate specified the user application. Moreover, every node has little information about the global parameter, as it has no idea about what is happening beyond its parent. In proposed approach, a node accepts estimations from all of its neighbors, and gradually gains in knowledge about the global information. It helps a node to understand whether its own information is useful to its neighbors. If a node realizes that its estimate could be useful to its neighbors, it transmits the new estimate. Unlike snapshot aggregation where the node transmits its estimate to its parent, in the proposed scheme, the node broadcasts its estimate to all its neighbors. Moreover, there is no need to establish and maintain a hierarchical relationship among the nodes in the network. This makes the algorithm particularly suitable for multiple user, mobile users, faulty nodes and transient network partition situations.

The proposed algorithm has the following steps:

1. Every node has an estimate of the global aggregated value (global estimate) in the form of (mean, covariance matrix). When a node makes a new local measurement, it makes an aggregation of the local observation with its current estimate. This is depicted in the block *Data Aggregation 1* in Fig. 5. The node computes the new global estimate and decides whether it should broadcast the new estimate to its neighbors. The decision is based on a threshold value as explained in Section 8.4.
2. When a node receives a global estimate from a neighbor, it first checks whether the newly received estimate differs from its current estimate by more than a pre-defined threshold.
    a. If the difference does not exceed the threshold, the node makes an aggregation of the global estimates (its current value and the received value) and computes a new global estimate. This is depicted in the block *Data Aggregation 2* in Fig. 5. The node then decides whether it should broadcast the new estimate.

    b. If the difference exceeds the threshold, the node performs the same function as in step (a). Additionally, it requests its other neighbors to send their values of the global estimate.
    c. If the estimates sent by the majority of the neighbors differ from the estimate sent by the first neighbor by a threshold value, then the node is assumed to be compromised. Otherwise, it is assumed to be normal.
3. If a node is identified to be compromised, the global estimate previously sent by it is ignored in the computation of the new global estimate and the node is isolated from the network by a broadcast message in its neighborhood.

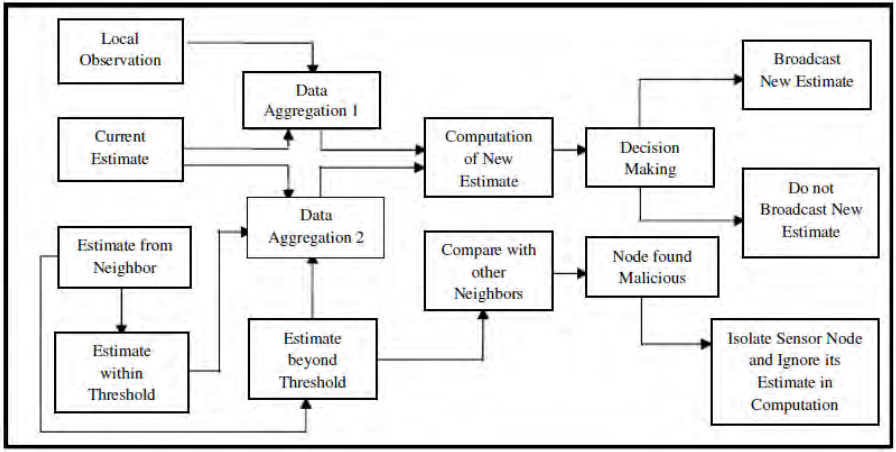

Fig. 5. A Schematic flow diagram of the proposed aggregation algorithm

**8.2 Aggregation of two global estimates**

In Fig. 5, the block *Data Aggregation 1* corresponds to this activity. For combining two global estimates to produce a single estimate, *covariance intersection* (CI) algorithm is used. CI algorithm is particularly suitable for this purpose, since it has the capability of aggregating two estimates without requiring any prior knowledge about their degree of correlation. This is more pertinent to WSNs, as we cannot guarantee statistical independence of observed data in such networks.

Given two estimates (*A*, $P_{AA}$) and (*B*, $P_{BB}$), the combined estimate (*C*, $P_{CC}$) by CI is given by (23) and (24):

$$P_{CC} = (\omega * P_{AA}^{-1} + (1-\omega) P_{BB}^{-1})^{-1} \qquad (23)$$

$$C = P_{CC}(\omega * P_{AA}^{-1} * A + (1-\omega) P_{BB}^{-1} * B) \qquad (24)$$

Here, $P_{AA}$, $P_{BB}$, and $P_{CC}$ represent the covariance matrices associated with the estimates *A*, *B*, and *C* respectively. The main computational problem with CI is the computation of $\omega$. The value of $\omega$ lies between 0 and 1. The optimum value of $\omega$ is arrived at when the trace of the determinant of $P_{CC}$ is minimized.

For *max* aggregation function, covariance matrices are simple scalars. It can be observed from (23) and (24) that in such a case $\omega$ can be either 1 or 0. Subsequently, $P_{CC}$ is equal to the minimum of $P_{AA}$ and $P_{BB}$, and $C$ is equal to either $A$ or $B$ depending on the value of $P_{CC}$. Even when the estimates are reasonably small-sized vectors, there are efficient algorithms to determine $\omega$.

### 8.3 Aggregation of a local observation with a global estimate

This module corresponds to the block *Data Aggregation 2* in Fig. 5. Aggregation of a local observation with a global estimate involves a statistical computation with two probability distributions.

**Case 1:** Mean of the local observation is greater than the mean of the current global estimate: In case of *max* aggregation function, if the mean of the local observation is greater than the mean of the current global estimate, the local observation is taken as the new estimate. The distribution of the new estimate is arrived at by multiplying the distribution of the current global estimate by a positive fraction ($w_1$) and summing it with the distribution of the local observation. The fractional value determines the relative weight assigned to the value of the global estimate. The weight assigned to the local observation being unity.

**Case 2**: Mean of the local observation is smaller than the mean of the current global estimate: If a node observes that the mean of the local observation is smaller than its current estimate, it combines the two distributions in the same way as in *Case 1* above, but this time a higher weight ($w_2$) is assigned to the distribution having the higher mean (i.e. the current estimate). However, as observed in (Boulis et al., 2003), this case should be handled more carefully if there is a sharp fall in the value of the global maximum. We follow the same approach as proposed in (Boulis et al., 2003). If the previous local measurement does not differ from the global estimate beyond a threshold value, a larger weight is assigned to the local measurement as in *Case 1*. In this case, it is believed that the specific local measurement is still the global aggregated value.

For computation of the weights $w_1$ and $w_2$ in *Case 1* and *Case 2* respectively, we follow the same approach as suggested in (Boulis et al., 2003). Since all the local measurements and the global estimates are assumed to follow Gaussian distribution, almost all the observations are bounded within the interval $[\mu \pm 3*\sigma]$. When the mean of the local measurement is larger than the mean of the global estimate, the computation of the weight ($w_1$) is done as follows. Let us suppose that $l(x)$ and $g(x)$ are the probability distributions for the local measurement and the global estimate respectively. If $l(x)$ and $g(x)$ can take non-zero values in the intervals $[x_1, x_2]$ and $[y_1, y_2]$ respectively, then the weight $w_1(x)$ will be assigned a value of 0 for all $x \leq \mu_1 - 3*\sigma$ and $w_1(x)$ will be assigned a value of 1 for all $x > \mu_1 - 3*\sigma$. Here, $x_1$ is equal to $\mu_1 - 3*\sigma_1$, where $\mu_1$ and $\sigma_1$ are the mean and the standard deviation of $l(x)$ respectively.

When the mean of the local measurement is smaller than the mean of the global estimate, the computation of the weight $w_2$ is carried out as follows. The value of $w_2(x)$ is assigned to be 0 for all $x \leq max\{\mu_1 - 3*\sigma_1, \mu_2 - 3*\sigma_2\}$. $w_2(x)$ is assigned a value of 1 for all $x > max\{\mu_1 - 3*\sigma_1, \mu_2 - 3*\sigma_2\}$. Here, $y_1$ is equal to $\mu_2 - 3*\sigma_2$, where $\mu_2$ and $\sigma_2$ represent the mean and the standard deviation of $g(x)$ respectively.

In all these computations, it assumed that the resultant distribution after combination of two bounded Gaussian distributions is also a Gaussian distribution. This is done in order to maintain the consistency of the estimates. The mean and the variance of the new Gaussian distribution represent the new estimate and the confidence (or certainty) associated with this new estimate respectively.

### 8.4 Optimization of communication overhead

Optimization of communication overhead is of prime importance in resource constrained and bandwidth-limited WSNs. The block named *Decision Making* in Fig. 5 is involved in this optimization mechanism of the proposed scheme. This module makes a trade-off between energy requirement and accuracy of the aggregated results.

To reduce the communication overhead, each node in the network communicates its computed estimate only when the estimate can bring a significant change in the estimates of its neighbors. For this purpose, each node stores the most recent value of the estimate it has received from each of its neighbors in a table. Every time a node computes its new estimate, it checks the difference between its newly computed estimate with the estimates of each of its neighbors. If this difference exceeds a pre-set threshold for any of its neighbors, the node broadcasts its newly computed estimate. The determination of this threshold is crucial as it has a direct impact on the level of accuracy in the global estimate and the energy expenditure in the WSN. A higher overhead due to message broadcast is optimized by maintaining two-hop neighborhood information in each node in the network (Boulis et al., 2003). This eliminates communication of redundant messages. This is illustrated in the following example.

Suppose that nodes *A*, *B* and *C* are in the neighborhood of each other in a WSN. Let us assume that node *A* makes a local measurement and this changes its global estimate. After combining this estimate with the other estimates of its neighbors as maintained in its local table, node *A* decides to broadcast its new estimate. As node *A* broadcasts its computed global estimate, it is received by both nodes *B* and *C*. If this broadcast estimate changes the global estimate of node *B* too, then it will further broadcast the estimate to node *C*, as node *B* is unaware that the broadcast has changed the global estimate of node *C* also. Thus the same information is propagated in the same set of nodes in the network leading to a high communication overhead in the network.

To avoid this message overhead, every node in the network maintains its two-hop neighborhood information. When a node receives information from another node, it not only checks the estimate values of its immediate neighbors as maintained in its table but also it does the same for its two-hop neighbors. Thus in the above example, when node *B* receives information from node *A*, it does not broadcast as it understands that node *C* has also received the same information from node *A*, since node *C* is also a neighbor of node *A*. The two-hop neighborhood information can be collected and maintained by using algorithms as proposed in (McGlynn & Borbash, 2001).

The choice of the threshold value is vital to arrive at an effective trade-off between the energy consumed for computation and the accuracy of the result of aggregation. For a proper estimation of the threshold value, some idea about the degree of dynamism of the physical process being monitored is required. A more dynamic physical process puts a greater load on the estimation algorithm thereby demanding more energy for the same level of accuracy (Boulis et al., 2003). If the user has no information about the physical process, he can determine the level of accuracy of the aggregation and the amount of energy spent dynamically as the process executes.

### 8.5 Security in aggregation scheme

The security module of the proposed scheme assumes that the sensing results for a set of sensors in the same neighborhood follows a normal (Gaussian) distribution. Thus, if a node receives estimates from one (or more) of its neighbors that deviates from its own local estimate by more than three times its standard deviation, then the neighbor node is suspected to have been compromised or failed. In such a scenario, the node that first detected such an anomaly sends a broadcast message to each of its neighbors requesting for the values of their estimates. If the sensing result of the suspected node deviates significantly (i.e., by more than three times the standard deviation) from the observation of the majority of the neighbor nodes, then the suspected node is detected as malicious. Once a node is identified as malicious, a broadcast message is sent in the neighborhood of the node that detected the malicious node and the suspected node is isolated from the network activities.

However, if the observation of the node does not deviate significantly from the observations made by the majority of its neighbors, the suspected node is assumed to be not malicious. In such a case, the estimate sent by the node is incorporated in the computation of the new estimate and a new global estimate is computed in the neighborhood of the node.

## 9. Simulation results

In this section, we describe the simulations that have been performed on the proposed scheme. As the proposed algorithm is an extension of the algorithm presented in (Boulis et al., 2003), we present here the results that are more relevant to our contribution, i.e., the performance of the security module. The results related to the energy consumption of nodes and aggregation accuracy for different threshold values (discussed in Section 8.4) are presented in detail in (Boulis et al., 2003) and therefore these are not within the scope of this work.

In the simulated environment, the implemented application accomplishes temperature monitoring, based on network simulator (*ns-2*) and its sensor network extension *Mannasim* (Mannasim, 2002). The nodes sense the temperature continuously and send the maximum sensed temperature only when it differs from the last data sent by more than 2%.In order to simulate the temperature behaviour of the environment, random numbers are generated following a Gaussian distribution, taking into consideration standard deviation of 1°C from an average temperature of 25°C. The simulation parameters are presented in Table 3.

To evaluate the performance of the security module of the proposed algorithm, two different scenarios are simulated. In the first case, the aggregation algorithm is executed in the nodes without invoking the security module to estimate the energy consumption of the aggregation algorithm. In the second case, the security module is invoked in the nodes and some of the nodes in the network are intentionally compromised. This experiment allows us to estimate the overhead associated with the security module of the algorithm and its detection effectiveness.

| Parameter | Value |
|---|---|
| No. of nodes | 160 |
| Simulation time | 200 s |
| Coverage area | 120 m * 120 m |
| Initial energy in each node | 5 Joules |
| MAC protocol | IEEE 802.11 |
| Routing protocol | None |
| Node distribution | Uniform random |
| Transmission power of each node | 12 mW |
| Transmission range | 15 m |
| Node capacity | 5 buffers |
| Energy spent in transmission | 0.75 W |
| Energy spent in reception | 0.25 mW |
| Energy spent in sensing | 10 mW |
| Sampling period | 0.5 s |
| Node mobility | Stationary |

Table 3. Simulation parameters

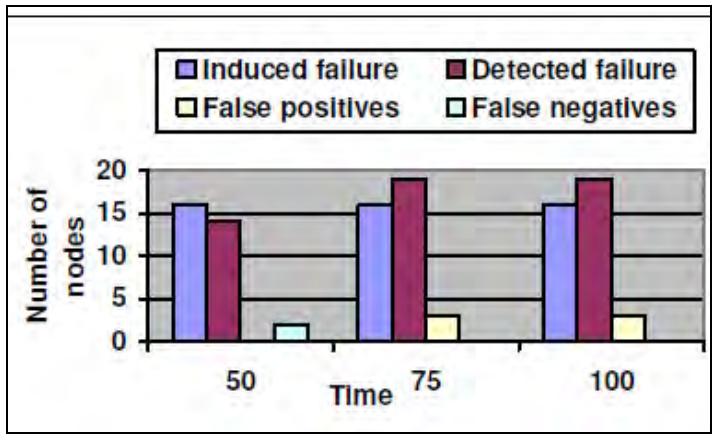

Fig. 6. Detection effectiveness with 10% of the nodes in the network faulty

It is observed that *delivery ratio* (ratio of the packets sent to the packets received by the nodes) is not affected by invocation of the security module. This is expected, as the packets are transmitted in the same wireless environment, introduction of the security module should not have any influence on the delivery ratio.

Regarding energy consumption, it is observed that the introduction of the security module has introduced an average increase of 105.4% energy consumption in the nodes in the network. This increase is observed when 20% of the nodes chosen randomly are compromised intentionally when the aggregation algorithm was executing. This increase in energy consumption is due to additional transmission and reception of messages after the security module is invoked.

To evaluate the detection effectiveness of the security scheme, further experiments are conducted. For this purpose, different percentage of nodes in the network is compromised and the detection effectiveness of the security scheme is evaluated. Fig. 6 and Fig. 7 present the results for 10% and 20% compromised node in the network respectively. In these diagrams, the false positives refer to the cases where the security scheme wrongly identifies a sensor node as faulty while it is actually not so. False negatives, on the other hand, are the cases where the detection scheme fails to identify a sensor node which is actually faulty. It is observed that even when there are 20% compromised nodes in the network the scheme has a very high detection rate with very low false positive and false negative rate. The results show that the proposed mechanism is quite effective in detection of failed and compromised nodes in the network.

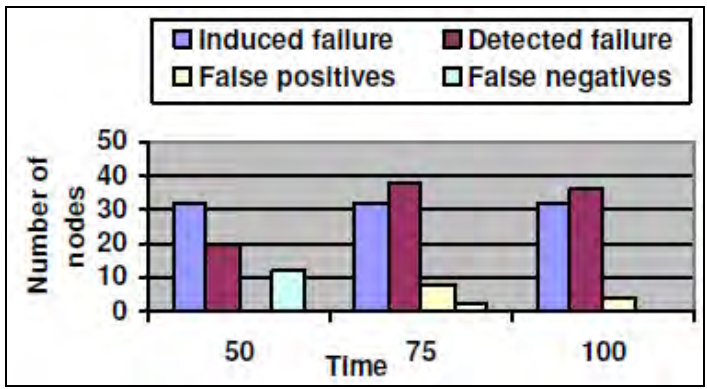

Fig. 7. Detection effectiveness with 20% of the nodes in the network faulty

## 10. Conclusion and future research issues

In-network data aggregation in WSNs is a technique that combines partial results at the intermediate nodes en route to the base station (i.e. the node issuing the query), thereby reducing the communication overhead and optimizing the bandwidth utilization in the wireless links. However, this technique raises privacy and security issues of the sensor nodes which need to share their data with the aggregator node. In applications such as health care and military surveillance where the sensitivity of the private data of the sensors is very high, the aggregation has to be carried out in a privacy-preserving way, so that the sensitive data are not revealed to the aggregator. A very popular scheme for this purpose exists in the literature which is known as CPDA. Although CPDA is in literature for quite some time now, no vulnerability of the protocol has been identified so far. In this chapter, we have first demonstrated a security vulnerability in the CPDA protocol, wherein a malicious sensor node can exploit the protocol is such a way that it gets access to the private values of its neighbors while participating in data aggregation process. A suitable modification of the CPDA protocol is further proposed so as to plug the identified vulnerability and also to make the protocol computationally more efficient. We have also made an analysis of the communication and computational overhead in the original CPDA protocol and the proposed modified version of the CPDA protocol. It has been found from the analysis that the modified version of the protocol involves appreciably less message communication overhead in the network and computational load on the sensor nodes.

We have also presented a comprehensive discussion on the existing secure aggregation protocols for WSNs and proposed a secure aggregation protocol for defending against attacks by malicious insider nodes that may introduce fake messages/data or alter data of honest nodes in the network. The performance of the proposed scheme has been evaluated on a network simulator and results have shown that the scheme is effective for defending attacks launched by malicious insider nodes in a WSN.

It may be noted that over the past few years, several schemes have been proposed in the literature for privacy preserving data aggregation in WSNs. A very popular and elegant approach in this direction is *homomorphic encryption* (Fontaine & Galand, 2007). Westhoff et al. have proposed additive privacy homomorphic functions that allow for end-to-end encryption between the sensors and the sink node and simultaneously enable aggregators to apply aggregation functions directly over the ciphertexts (Westhoff et al., 2006). This has the advantage of eliminating the need for intermediate aggregators to carry out decryption and encryption operations on the sensitive data. Armknecht et al. have presented a symmetric encryption scheme for sensor data aggregation that is homomorphic both for data and the keys (Armknecht et al., 2008). This is called *bi-homomorphic encryption*, which is also essentially an additive homomorphic function. Castellucia et al. have proposed an approach that combines inexpensive encryption techniques with simple aggregation methods to achieve efficient aggregation of encrypted data in WSNs (Castelluccia et al., 2009). The method relies on end-to-end encryption of data and hop-by-hop authentication of nodes. Privacy is achieved by using additive homomorphic functions. A very simple approach for privacy-preserving multi-party computation has been discussed by Chaum (Chaum, 1988). The protocol is known as *Dining Cryptographers Problem* which describes the way a channel is created so that it is difficult to trace (i.e. identify) the sender of any message through that channel.

The approaches based on privacy homomorphic functions are more elegant than CPDA for the purpose of carrying out sensor data aggregation in a privacy preserving way. However, they involve large computational overhead due to complexities involved in computing the homomorphic encryption functions and the associated key management related issues. Most of the existing public key cryptography-based privacy homomorphic functions are too heavy for resource-constrained battery-operated sensor nodes. Some secure data aggregation schemes use elliptic curve cryptography (Westhoff et al., 2006). However, these schemes work only for some specific query-based aggregation functions, e.g., sum, average etc. A more elegant scheme that works for all types of functions is clearly in demand. In (Gentry, 2009), a fully homomorphic function has been presented. However, this scheme is too complex and heavy-weight for deployment in WSNs. In addition, in some WSN environment, symmetric cryptography-based privacy homomorphic encryption schemes are more suitable (Castelluccia, 2005; Castelluccia, 2009; Ozdemir, 2008). However, most of the current homomorphic encryption schemes are based on public key encryption. Hence, exploration of symmetric key cryptography based privacy homomorphism functions is an interesting research problem. Another emerging research problem is the use of *digital watermarking* schemes in place of privacy homomorphic encryption functions (Zhang et al., 2008). However, this method allows only one-way authentication of sensor data at the base station only. To defend against *rogue base station attacks* on sensor nodes, this scheme would not be applicable. Design of mutual authentication scheme using watermarking techniques for secure and privacy-preserving data aggregation protocols is another research problem that needs attention of the research community.